\documentclass[a4paper,11pt]{article}
\usepackage{jheppub} 
\usepackage{lineno}
\usepackage{xcolor}
\usepackage{amssymb}
\usepackage{physics}
\graphicspath{{images/}}

\title{\boldmath  Quantum-information diagnostics of cosmological perturbations with nontrivial sound speed in inflation} 

\author[a]{Shi-Cheng Liu, }
\author[a] { Lei-Hua Liu,  }
\author[a] { Bichu Li, }
\affiliation[a]{Department of Physics, College of Physics, Mechanical and Electrical Engineering, Jishou University, Jishou 416000, China }

\author[b,c]{Hai-Qing Zhang, }
\affiliation[b]{Center for Gravitational Physics, Department of Space Science, Beihang University, Beijing, 100191, China }
\affiliation[c]{Peng Huanwu Collaborative Center for Research and Education, Beihang University, Beijing
100191, China }
\author[d] { Peng-Zhang He  }
\affiliation[d]{School of Physics and Astronomy, China West Normal University, Nanchong 637002, Sichuan, China}

\emailAdd{2023700328@stu.jsu.edu.cn, liuleihua8899@hotmail.com (corresponding author), libichu@mail.ustc.edu.cn, hqzhang@buaa.edu.cn, hepzh@cwnu.edu.cn}

\abstract{In this work, we systematically investigate the quantum-information diagnostics of cosmological perturbations with a nontrivial sound speed, utilizing a normalized open two-mode squeezed-state framework. Rather than introducing new observables, our analysis focuses on how a modified sound speed dynamically reshapes the Schrödinger evolution of the squeezing parameters ($r_k$ and $\phi_k$). We demonstrate how these dynamical changes are inherited by the reduced density matrix of the observable sector. By employing a sound-speed-resonance parametrization, we derive and evaluate the purity, von Neumann entropy, Rényi entropies, and logarithmic negativity. We introduce a bounded variable $x = \tanh r_k$ and perform numerical simulations within the inflationary regime. Our numerical results reveal that a nontrivial sound speed significantly suppresses the purity of the reduced state, indicating enhanced effective mixedness. Simultaneously, it strongly amplifies and modulates both the entropic and entanglement diagnostics. More precisely, a nontrivial sound speed postpones the onset of classicality by modulating the decoherence process. Ultimately, we show that a nontrivial sound speed leaves distinct and identifiable quantum-information signatures within the entanglement structure of the early universe.
}

\begin{document}
\maketitle
\flushbottom
\section{Introduction}
\label{sec:intro}

The quantum origin of primordial cosmological perturbations is of central importance in modern cosmology. While the nature of the initial state remains a subject of active research, the Bunch–Davies (BD) vacuum serves as a robust candidate for exploring the Cosmic Microwave Background (CMB) \cite{Agullo:2022ttg}. In curved spacetime, defining a unique vacuum state is generally non-trivial \cite{Unruh:1976db}, except in cases of constant curvature, such as de Sitter space. During inflation, vacuum fluctuations are amplified and evolve into highly squeezed two-mode states, making the squeezed-state formalism the natural language for describing the $(\vec{k},-\vec{k})$ sector of cosmological perturbations \cite{Grishchuk:1990bj,albrecht1994inflation}. This perspective has long been instrumental in understanding the quantum-to-classical transition of primordial fluctuations and the emergence of effectively classical stochastic perturbations on super-Hubble scales \cite{Polarski:1995jg,Mukhanov:1981xt,Kiefer:2008ku}. Furthermore, even the quantum perturbations generated during inflation are expected to undergo rapid decoherence into classicality when accounting for environmental interactions \cite{Burgess:2015ajz}.

A particularly compelling departure from the canonical single-field paradigm occurs when scalar perturbations propagate with a non-trivial sound speed, $c_s \neq 1$ \cite{Armendariz-Picon:1999hyi,Alishahiha:2004eh,Cheung:2007st,Peiris:2007gz,Garriga:1999vw}. Such scenarios arise naturally in non-canonical inflationary models—including $k$-inflation, DBI inflation, and Effective Field Theory (EFT) frameworks—where heavy degrees of freedom have been integrated out. Because the sound speed enters the perturbation Hamiltonian directly, it is expected to modify not only the observable power spectra but also the dynamical squeezing history of the underlying quantum state. To analyze these effects, we adopt a sound-speed-resonance (SSR) parametrization \cite{Cai:2018tuh} as an effective means of encoding oscillatory sound-speed modulations into the perturbation dynamics. Simultaneously, the study of cosmological perturbations has been significantly enriched by insights from quantum information theory \cite{Brahma:2020zpk,Shandera:2017qkg,Burgess:2025dwm,Cespedes:2025zqp,Kharel:2025lek,Salcedo:2025ezu,Belfiglio:2025cst,Burgess:2024eng,Colas:2024xjy,Boutivas:2023mfg,Ning:2023ybc,Boutivas:2023ksg,Giantsos:2022qdd,Rai:2020edx,Szapudi:2002cr,Martin:2007bw,Bhargava:2020fhl,Bardeen:1986iq}. Rather than assuming a globally pure state, recent developments favor an open-system perspective, in which the observable perturbation sector is treated as a subsystem coupled to an environment \cite{Bhattacharyya:2024duw}. Consequently, the system is more appropriately described by a reduced density matrix (RDM) and an open two-mode squeezed state (OTMSS) framework \cite{Liu:2025caj,Zhai:2025abc,Zhai:2024tkz,Zhai:2024odw,Li:2024kfm}, which can be further mapped to the double thermal field formalism \cite{Li:2024ljz,Li:2024iji}. Information-theoretic observables—such as entanglement entropy, purity, and logarithmic negativity—offer a more refined diagnostic of the quantum correlations and effective mixedness of inflationary states than the conventional pure-state language.

Despite these parallel advances, a significant gap remains in our systematic understanding of the early universe. On the one hand, investigations into non-trivial sound speeds have predominantly focused on macroscopic and observational signatures, such as modifications to the primordial power spectrum, the generation of large non-Gaussianities, and the enhanced production of primordial black holes (PBHs) \cite{Chen:2006nt,Achucarro:2010da,Pi:2022ysn,Byrnes:2018txb,Miranda:2012rm,Garcia-Bellido:2017mdw}. On the other hand, recent quantum-information analyses of cosmological perturbations—utilizing diagnostic tools like entanglement entropy, quantum discord, and decoherence bounds—have been carried out almost exclusively within the canonical framework ($c_s = 1$) \cite{Martin:2015qta,Nambu:2008my,Kanno:2017dci,Lim:2014uea,Esteban:2022rjk}, or have adopted open-system descriptions without explicitly incorporating modified kinetic dynamics \cite{Nelson:2016kjm,Hollowood:2017bil,Kiefer:2006je}.What is notably absent in the literature is a systematic, reduced-state analysis that unifies an open two-mode squeezed state (OTMSS) description with a non-trivial sound speed. Specifically, it remains unexplored how the modified Schrödinger evolution of the squeezing parameters, driven by $c_s \neq 1$, dynamically reshapes standard quantum-information diagnostics. The present work aims to bridge this gap.

Several recent studies have utilized circuit complexity to investigate the properties of the early universe \cite{Bhattacharyya:2020rpy,Bhattacharyya:2020kgu}, including our own previous work exploring inflation via circuit complexity with non-kinetic modifications \cite{Liu:2021nzx,Li:2021kfq,Li:2023ekd}. In the present work, we systematically investigate the quantum-information diagnostics of cosmological perturbations within a normalized OTMSS framework \cite{Liu:2025caj,Zhai:2025abc,Zhai:2024tkz,Zhai:2024odw,Li:2024kfm}. The central premise of our analysis is that a modified sound speed does not merely introduce new algebraic definitions for information-theoretic observables. Instead, it enters the perturbation Hamiltonian directly, thereby driving a modified Schrödinger evolution for the squeezing amplitude $r_k$ and phase $\phi_k$.By tracing out one momentum sector, we construct the reduced density matrix (RDM) for the observable mode, from which we derive the purity, von Neumann entropy, Rényi entropies, and logarithmic negativity. Consequently, the physical imprints of the sound-speed resonance are dynamically inherited by these reduced-state diagnostics. 
We reformulate the dynamics in terms of the bounded variable $x = \tanh r_k$ \cite{Polarski:1995jg}. While our theoretical framework is formally valid across subsequent cosmological epochs, our numerical analysis focuses on the inflationary regime.

Through a systematic investigation of quantum-information diagnostics, we find that sound-speed resonance significantly modifies the oscillatory evolution of both the squeezing amplitude and phase. These dynamical modifications are intrinsically inherited by the reduced-state measures: compared to the canonical scenario ($c_s=1$), the purity of the reduced state is notably suppressed, indicating enhanced effective subsystem mixedness. Simultaneously, the von Neumann entropy, Rényi entropies, and logarithmic negativity are enhanced and exhibit pronounced oscillatory modulations \cite{Maldacena:2012xp}. Collectively, these results demonstrate that a non-trivial sound speed does more than merely alter macroscopic power spectra; it leaves distinct, identifiable quantum-information signatures within the entanglement structure of the early universe.

The remainder of this paper is organized as follows. In Sec.~\ref{Cosmological background and nontrivial sound speed}, we will introduce the cosmological background and the non-trivial sound-speed parametrization. In Sec.~\ref{Open-system squeezed-state framework}, we will present the normalized OTMSS framework and derive the governing evolution equations for the squeezing variables. In Sec.~\ref{Reduced density matrix and quantum-information diagnostics}, we will construct the RDM and define the specific quantum-information diagnostics for the OTMSS. Our numerical results for the inflationary regime, along with a discussion of their physical implications, are presented in Sec.~\ref{Numerical results in the inflationary regime}. Finally, we will offer concluding remarks and an outlook on future research in Sec.~\ref{Conclusions}.

\section{Some set up for inflation and non-trivial sound speed}
\label{Cosmological background and nontrivial sound speed}
In this section, we will give some foundation of the scale factor in inflation, and meanwhile we will introduce the SSR as the non-trivial sound speed \cite{Cai:2018tuh}.

\subsection{Background evolution in inflation}
In this work, we consider a spatially flat Friedmann--Lema\^itre--Robertson--Walker (FLRW) metric as a reasonable approximation, 
\begin{equation}
ds^2 = a^2(\eta)\left(-d\eta^2 + d\vec{x}^{\,2}\right),
\label{FLRW metric}
\end{equation}
where $a(\eta)$ is the scale factor in terms of  conformal time and $\vec{x}$ is the spatial part. 
For simplicity, we will follow the notation of \cite{Baumann:2009ds} to define single component universe, where the equation of state parameter (EoS) is 
\begin{equation}
w_I = \frac{P_I}{\rho_I},
\end{equation}
where $P_I$ and $\rho_I$ are the pressure and energy density in the epoch $I$ including inflation, RD and MD.
In this work, we will focus on inflation.

To relate the scale factor to conformal time, we first define the comoving particle horizon, 
\begin{equation}
    \chi_{\rm ph}=\int_{t_i}^t\frac{dt'}{a(t')}=\int_{\ln a_i}^{\ln a} d\ln a',
    \label{comving distance}
\end{equation}
where $\chi_{\rm ph}$ denotes the comoving distance of particle horizon. The Hubble radius $(aH)^{-1}$ is represented by the Eos as follows,
\begin{equation}
(aH)^{-1} = H_0^{-1} a^{\frac{1}{2}(1+3w)}.
\end{equation}
Substituting this relation into the definition of conformal time and integrating yields
\begin{equation}
\eta = \frac{2H_0^{-1}}{1+3w} a^{\frac{1}{2}(1+3w)} + C,
\end{equation}
where $C$ is an integration constant fixed by boundary conditions and we can set safely set $C=0$. In inflation, we have $w=-1$ and then could derive the conformal time for inflation 
\begin{equation}
    \eta=-(aH_0)^{-1}
\end{equation}
where $H_0$ is current value for the Hubble parameter. 
The FLRW metric \eqref{FLRW metric} and conformal time will be adopted to the following calculations.

\subsection{Nontrivial sound speed and sound-speed resonance}
\label{Nontrivial sound speed and sound-speed resonance}

The second essential ingredient in our setup is a nontrivial sound speed
$c_s$, which modifies the propagation of scalar cosmological perturbations.
In the canonical single-field slow-roll scenario one has $c_s^2=1$, whereas
$c_s^2\neq 1$ usually signals departures from the canonical kinetic structure.
Such a situation naturally appears in noncanonical inflationary models,
including $k$-inflation, DBI inflation, and effective field theory descriptions
of inflation, where heavy degrees of freedom can generate an effective reduced
sound speed for the adiabatic perturbation sector.

In the present work, we adopt the SSR mechanism as a
phenomenological parametrization of an oscillatory sound speed \cite{Cai:2018tuh}. The basic idea
of SSR is that a periodically varying sound speed can induce a parametric
resonance in the Mukhanov--Sasaki equation in inflation. As a consequence, the curvature perturbation can be strongly amplified around characteristic scales. This mechanism was originally proposed as an efficient way to enhance primordial density perturbations and to produce primordial black holes during inflation. It was subsequently studied in the inflaton--curvaton
mixed scenario \cite{Chen:2019zza}, in explicit DBI realizations \cite{Chen:2020uhe}, and in analyses of backreaction effects associated with the enhanced scalar perturbations \cite{Li:2023zva}. More recently, related
sound-speed-resonance effects have also been discussed in the tensor sector and in modified-gravity or Gauss--Bonnet-coupled inflationary models \cite{Addazi:2024gew}.
Following the standard SSR ansatz \cite{Cai:2018tuh}, we take
\begin{equation}
    c_s^2(\eta)=1-2\xi\left[1-\cos(k\eta)\right],
    \label{eq:ssr_ansatz}
\end{equation}
where $\xi$ denotes the oscillation amplitude and $k$ is the comoving momentum mode. The canonical limit is recovered when $\xi=0$, for which
$c_s^2=1$. For $\xi\neq 0$, the oscillatory part of $c_s^2$ modifies the
effective frequency of the perturbation mode and therefore changes the squeezing dynamics generated by the perturbation Hamiltonian. Since the present numerical analysis focuses on the inflationary regime, we only need the inflationary form of the SSR profile, where we use $\eta=-\frac{1}{aH_0}$ and introduce $y=\log_{10}a$, the SSR parametrization becomes
\begin{equation}
    c_s^2(y)=1-2\xi\left[1-\cos\left(k\,10^{-y}\right)\right].
    \label{eq:ssr_inflation_y}
\end{equation}
This expression will be used throughout the inflationary numerical analysis. In our framework, the role of $c_s$ is not to redefine the quantum-information measures themselves. Instead, $c_s$ enters the Hamiltonian explicitly and therefore modifies the Schr\"odinger evolution of the squeezing amplitude
$r_k$ and squeezing phase $\phi_k$. The influence of SSR is then inherited dynamically by the RDM, the purity, the von Neumann entropy,
the R\'enyi entropies, and the logarithmic negativity.

\subsection{SSR in inflation}
\label{SSR in inflation}

Having specified the inflationary background and the SSR parametrization, we now clarify the precise role played by the nontrivial sound speed in the present work. The key point is that
$c_s \neq 1$ does not alter the formal definitions of the quantum-information observables themselves. Instead, its effect is dynamical: the sound speed enters the quadratic Hamiltonian of cosmological perturbations explicitly and
thereby modifies the Schr\"odinger evolution of the open two-mode squeezed state. As a result, the squeezing amplitude $r_k$ and the squeezing phase
$\phi_k$ acquire a nontrivial $c_s$-dependence, and this modified squeezing history is subsequently inherited by the RDM and all derived diagnostics.
This observation is central to the logic of the paper. Quantities such as the purity, the von Neumann entropy, the R\'enyi entropies, and the logarithmic
negativity are not redefined in the presence of SSR. Their mathematical forms remain the standard reduced-state expressions. However, since these quantities
are ultimately determined by the squeezing data of the mode pair $r_k,~\phi_k$, any change in the sound speed affects them indirectly but physically through the underlying time evolution. In this sense, the imprint of nontrivial sound
speed is encoded not in a new algebraic structure of the observables, but in the dynamical trajectory of the quantum state from which those observables are
constructed.

At the formal level, the framework developed in this paper can be written for different cosmological epochs once the background evolution and the corresponding sound-speed profile are specified. For the numerical analysis, we rewrite the squeezing dynamics in terms of the bounded variable
\begin{equation}
    x \equiv \tanh r_k,
\end{equation}
and focus on the inflationary interval
\begin{equation}
    -1 \le y=\log_{10}a \le 0,
\end{equation}
which is sufficient for the main purpose of the work, namely to isolate how the SSR-induced modulation of the sound speed reshapes the squeezing dynamics during inflation and how these dynamical modifications are reflected in the reduced-state quantum-information diagnostics. Therefore, the strategy of the present analysis is clear: we first determine how the nontrivial sound speed modifies the evolution of $r_k$ and $\phi_k$ through the Schr\"odinger equation, and we then study how this modified
evolution is encoded in the purity, the entropic measures, and the logarithmic negativity. In this way, the physical effect of SSR is traced consistently from the perturbation Hamiltonian to the observable
quantum-information signatures of the inflationary state.

\section{Open-system squeezed-state framework}
\label{Open-system squeezed-state framework}
In this section, we will investigate the numeric of $r_k$ and $\phi_k$ in inflation under the framework of OTMSS, incorporating the effects of open system. First, we will show the generic two-mode Hamiltonian in terms of SSR.

\subsection{Hamiltonian for cosmological perturbations}

We now turn to the quantum description of cosmological perturbations in the
presence of a nontrivial sound speed. As emphasized in Sec.~\ref{Nontrivial sound speed and sound-speed resonance}, the role of
the SSR in the present work is dynamical rather than
algebraic: it does not redefine the quantum-information observables
themselves, but instead enters the perturbation Hamiltonian and thereby
modifies the Schr\"odinger evolution of the squeezed state. The construction
of the Hamiltonian is therefore the starting point for tracing how the
nontrivial sound speed affects the subsequent reduced-state diagnostics. We begin with the scalar sector of cosmological perturbations around the
spatially flat FLRW background. Introducing the curvature perturbation and
passing to the Mukhanov--Sasaki variable,
\begin{equation}
    f \equiv z \mathcal{R},
\end{equation}
with
\begin{equation}
    z \equiv \frac{\sqrt{2\epsilon}\,a}{c_s},
\end{equation}
the quadratic action for a single-field system with nontrivial sound speed
can be written in the schematic form
\begin{equation}
    S^{(2)}=\frac{1}{2}\int d\eta\, d^3x
    \left[
        f'^2
        - c_s^2 (\partial_i f)^2
        + \left(\frac{z'}{z}\right)^2 f^2
        - 2\frac{z'}{z} f'f
        - a^2 V_{\phi\phi} f^2
    \right],
    \label{eq:quad_action_f}
\end{equation}
where a prime denotes differentiation with respect to conformal time $\eta$, $\epsilon$ is the slow-roll parameter, and $V_{\phi\phi}$ is the second derivative of the inflaton potential. This form makes it explicit that the sound speed appears directly in the gradient term and therefore changes the effective mode frequency of the perturbation sector. From the action \eqref{eq:quad_action_f}, the canonical momentum conjugate to
$f$ is defined by
\begin{equation}
    \pi(\eta,\vec{x}) \equiv \frac{\partial \mathcal{L}}{\partial f'}.
\end{equation}
The corresponding Hamiltonian then takes the form
\begin{equation}
    H=\frac{1}{2}\int d^3x
    \left[
        \pi^2
        + c_s^2(\partial_i f)^2
        + \frac{z'}{z}(\pi f + f\pi)
        + a^2 V_{\phi\phi} f^2
    \right].
    \label{eq:hamiltonian_position}
\end{equation}
This expression is the natural starting point for quantization, since it makes transparent the two ingredients that are central to the present work:
the background-driven squeezing term proportional to $z'/z$, and the nontrivial sound-speed contribution proportional to $c_s^2$. We next Fourier-expand the field operator and its conjugate momentum in the
standard normalized mode basis,
\begin{equation}
    \hat{f}(\eta,\vec{x})
    =
    \int \frac{d^3k}{(2\pi)^{3/2}} \frac{1}{\sqrt{2k}}
    \left[
        \hat{c}_{\vec{k}}\, v_k(\eta)e^{i\vec{k}\cdot\vec{x}}
        + \hat{c}_{\vec{k}}^\dagger\, v_k^*(\eta)e^{-i\vec{k}\cdot\vec{x}}
    \right],
\end{equation}
\begin{equation}
    \hat{\pi}(\eta,\vec{x})
    =
    i\int \frac{d^3k}{(2\pi)^{3/2}} \sqrt{\frac{k}{2}}
    \left[
        \hat{c}_{\vec{k}}^\dagger\, u_k^*(\eta)e^{-i\vec{k}\cdot\vec{x}}
        - \hat{c}_{\vec{k}}\, u_k(\eta)e^{i\vec{k}\cdot\vec{x}}
    \right],
\end{equation}
where $\hat{c}_{\vec{k}}^\dagger$ and $\hat{c}_{\vec{k}}$ are the creation and annihilation operators. After imposing the normalization conditions on the mode functions and transforming to momentum space, the Hamiltonian for a given mode pair $(\vec{k},-\vec{k})$ can be written as
\begin{equation}
\begin{split}
\hat H_k =&
\left(
\frac{a^2V_{\phi\phi}}{2k}
+i\frac{z'}{z}
-\frac{k}{2}
+\frac{k c_s^2}{2}
\right)\hat c_{\vec k}^\dagger \hat c_{-\vec k}^\dagger
+
\left(
\frac{a^2V_{\phi\phi}}{2k}
-i\frac{z'}{z}
-\frac{k}{2}
+\frac{k c_s^2}{2}
\right)\hat c_{\vec k}\hat c_{-\vec k}
\\\\&+
\left(
\frac{a^2V_{\phi\phi}}{2k}
+\frac{k}{2}
+\frac{k c_s^2}{2}
\right)
\left(
\hat c_{\vec k}\hat c_{\vec k}^\dagger
+\hat c_{-\vec k}\hat c_{-\vec k}^\dagger
\right). 
\end{split}
\label{eq:Hk_final}
\end{equation}
This quadratic Hamiltonian is of the standard squeezed-state type. The terms proportional to $\hat{c}_{\vec{k}}^\dagger \hat{c}_{-\vec{k}}^\dagger$
and $\hat{c}_{\vec{k}} \hat{c}_{-\vec{k}}$
describe pair creation and pair annihilation of the mode pair $(\vec{k},-\vec{k})$, while the diagonal terms govern the phase rotation of the occupation-number sector. In particular, the coefficient $z'/z$ is responsible for the squeezing structure induced by the cosmological background, whereas the $kc_s^2/2$ contribution shows explicitly how the nontrivial sound speed modifies the effective mode dynamics. Equation~\eqref{eq:Hk_final} also makes clear why the influence of SSR is
ultimately inherited by the reduced-state quantum-information observables. The sound speed enters the Hamiltonian directly, the Hamiltonian determines
the Schr\"odinger evolution of the open two-mode squeezed state, and this evolution in turn determines the squeezing parameters $r_k$ and $\phi_k$.
Consequently, even though the purity, the von Neumann entropy, the R\'enyi entropy, and the logarithmic negativity retain their standard formal
definitions, they acquire nontrivial sound-speed dependence through the
dynamical history encoded in \eqref{eq:Hk_final}.

\subsection{The normalized OTMSS}
\label{normalized OTMSS}
The OTMSS was first derived in Ref. \cite{Zhai:2024odw} based primarily on the second Meixner polynomials. A key feature of the OTMSS is its model-independent nature; the specific dynamics of various models are encapsulated within two parameters: $u_2$ and $\sqrt{|1-u_1^2|}$. In this work, we provide the explicit formula for the OTMSS as follows:
\begin{equation}
    |\mathcal{O}(\eta)\rangle = \frac{\operatorname{sech} r_k(\eta)}{1 + u_2 \tanh r_k(\eta)} \sum_{n=0}^{\infty} |1 - u_1^2|^{\frac{n}{2}} \times \frac{(-\exp(2i\phi_k(\eta)) \tanh r_k(\eta))^n}{(1 + u_2 \tanh r_k(\eta))^n} |n, n\rangle_{-\vec{k}, \vec{k}},
    \label{OTMSS}
\end{equation}
where $u_2$ acts as a dissipation coefficient. As discussed in Refs. \cite{Zhai:2024odw, Li:2024kfm}, the numerical values of $u_2$ indicate that inflation behaves as a strongly dissipative system. The original OTMSS \eqref{OTMSS} is non-normalizable; however, normalization is required to satisfy the standard definitions of purity and other observables in quantum information theory. Following a straightforward algebraic procedure, the normalized OTMSS is derived as follows:
\begin{equation}
\begin{split}
|O(\eta)\rangle
=&
\frac{\sqrt{(1+u_2\tanh{r_k}(\eta))^2-|1-u_1^2|\tanh^2{r_k}(\eta)}}{1+u_2\tanh r_k(\eta)}
\\&\sum_{n=0}^{\infty}
|1-u_1^2|^{n/2}
\left[
-\frac{e^{2i\phi_k(\eta)}\tanh r_k(\eta)}{1+u_2\tanh r_k(\eta)}
\right]^n
|n,n\rangle_{-\vec k,\vec k},
\end{split}
\label{normalized otmss}
\end{equation}
where $r_k(\eta)$ and $\phi_k(\eta)$ represent the squeezing amplitude and squeezing phase, respectively, and the prefactor serves as the normalization constant. The detailed derivation of Eq. \eqref{normalized otmss} is provided in App. \ref{normalization of otmss}. Notably, the normalized OTMSS reduces to the standard two-mode squeezed state (TMSS) in the weak-dissipation limit $(u_2 \ll 1)$, 
\begin{equation}
|O(\eta)\rangle
=
\frac{1}{\cosh r_k}
\sum_{n=0}^{\infty}
(-1)^n e^{2in\phi_k}\tanh^n r_k\,|n,n\rangle_{\vec k,-\vec k}
+\mathcal{O}(u_2),
\label{two mode squeezed state}
\end{equation}
In the following calculations, we will derive the evolution equation of $r_k$ and $\phi_k$ according to the normalized OTMSS \eqref{normalized otmss}. 

\subsection{Schr\"odinger evolution of the squeezing parameters}
To obtain the evolution of $r_k$ and $\phi_k$ within the normalized OTMSS framework, we employ the Schrödinger equation to derive their respective governing equations, which are given by: 
\begin{equation}
\hat H\,|O(\eta)\rangle = i\,\partial_\eta |O(\eta)\rangle,
\end{equation}
where $\hat{H}$ is the Hamiltonian operator defined in Eq. \eqref{eq:Hk_final} in terms of conformal time. The resulting evolution equations for $r_k$ and $\phi_k$ are as follows:
\begin{equation}
    \begin{split}
        r'_k&=\frac{|1-u_1^2|^{\frac{1}{2}}[(1+u_2\tanh r_k)^2-|1-u_1^2|\tanh^2{r_k}][\frac{z'}{z}\cos{2\phi_k}-(\frac{a^2V_{\phi\phi}}{2k}-\frac{k}{2}+\frac{kc_s^2}{2})\sin{2\phi_k}]}{|1-u_1^2|\tanh^2{r_k}-|1-u_1^2|}\\&+\frac{\frac{1}{2}(1+u_2\tanh{r_k})|1-u_1^2|'\tanh{r_k}}{|1-u_1^2|\tanh^2{r_k}-|1-u_1^2|}\frac{u_2'|1-u_1^2|\tanh^2{r_k}}{|1-u_1^2|\tanh^2{r_k}-|1-u_1^2|}
    \end{split}
    \label{r prime}
\end{equation}
\begin{equation}
    \begin{split}
       \phi'_k=&-\frac{1}{2}(k+kc^2_s+\frac{a^2V_{\phi\phi}}{k})+\frac{1}{2}[(\frac{a^2V_{\phi\phi}}{2k}-\frac{k}{2}+\frac{kc^2_s}{2})\cos{2\phi_k}+\frac{z'}{z}\sin{2\phi_k}]\\&(|1-u_1^2|^{-\frac{1}{2}}(u_2+\coth{r_k})+|1-u_1^2|^{\frac{1}{2}}\frac{\tanh{r_k}}{1+u_2\tanh{r_k}})
    \end{split}
    \label{phik prime}
\end{equation}
where the prime denotes differentiation with respect to conformal time $\eta$. Their detailed derivation is provided in App. \ref{rk and phhi}. 
For convenience in the subsequent numerical analysis, we define the
\begin{equation}
y=\log_{10}a,
\label{y}
\end{equation}
for the later investigations. As discussed in Sec. \ref{Cosmological background and nontrivial sound speed}, the non-trivial sound speed $c_s$ is explicitly related to the scale factor. Consequently, we can utilize the variable $y$ to investigate the evolution of $r_k$ and $\phi_k$ throughout the entire history of the early universe, including the inflation, radiation-dominated (RD), and matter-dominated (MD) epochs. Furthermore, the oscillatory structure of $c_s$—specifically the sound-speed resonance (SSR) in our model—is more effectively visualized and analyzed through numerical simulations.

In principle, the entirety of the early universe can be investigated, as the scale factor can be characterized by the equation of state parameter $w$, which encodes the dynamics of different cosmological epochs; relevant studies using this approach can be found in our previous works \cite{Liu:2025caj, Zhai:2024odw, Li:2024ljz}. 
In the present analysis, we employ the bounded variable $x = \tanh r_k$ within the range $-1 \le y \le 0$ and focus on the inflationary period.

\section{RDM and quantum-information diagnostics}
\label{Reduced density matrix and quantum-information diagnostics}

To characterize the observable sector in the early Universe, it is useful to introduce the RDM by tracing out inaccessible or environmental degrees of freedom. If the total wave functional is denoted by $\Psi[\varphi,\chi;\eta]$, where $\varphi$ represents the relevant system variables (for example, long-wavelength cosmological perturbations) and $\chi$ denotes the environment, then the reduced density matrix is defined as
\begin{equation}
\rho_{\rm red}[\varphi,\tilde{\varphi};\eta]
=
\int {\cal D}\chi \,
\Psi[\varphi,\chi;\eta]\,
\Psi^{*}[\tilde{\varphi},\chi;\eta].
\end{equation}
Equivalently, in operator language one may write
\begin{equation}
\rho_{\rm red}(\eta)= {\rm Tr}_{E}\,\rho_{\rm tot}(\eta).
\end{equation}
The diagonal elements of $\rho_{\rm red}$ encode the probabilities of different field configurations, whereas the off-diagonal elements measure quantum coherence. Therefore, when the off-diagonal components are strongly suppressed, the subsystem behaves effectively as a classical statistical ensemble rather than as a pure quantum superposition. In this sense, the reduced density matrix provides a natural framework for discussing decoherence, entropy generation, and the quantum-to-classical transition of primordial fluctuations \cite{Brandenberger:1990bx,LaFlamme:1990kd,Paz:1991ze,Laflamme:1993zx,Calzetta:1995ys,Polarski:1995jg}.

In inflationary cosmology, the system is usually identified with the long-wavelength perturbations, while the environment may consist of short-wavelength modes, additional matter fields, or other inaccessible inhomogeneous degrees of freedom. Tracing out these environmental variables leads to a reduced description in which decoherence can dynamically select an approximately diagonal basis, often close to the field-amplitude basis. This makes it possible to interpret the primordial perturbations as classical stochastic seeds for later structure formation without abandoning their quantum origin \cite{Polarski:1995jg,Lesgourgues:1996jc,Kiefer:1998qe}, even considering the early universe as an open quantum system \cite{Lombardo:2005iz,Boyanovsky:2015tba,Nelson:2016kjm}. In the following investigations, we will utilize purity, von Neumann entropy, R\'enyi entropy, Logarithmic negativity to diagnose the decoherence for the cosmological perturbations in inflation.

\subsection{RDM of OTMSS}
\label{RDM of OTMSS}

Having established the open-system squeezed-state framework, we now turn to the quantum-information diagnostics associated with the mode pair $(\vec{k},-\vec{k})$.
The relevant object is the reduced density matrix obtained by tracing out one of the two modes.
Starting from the OTMSS \eqref{normalized otmss} introduced in Sec. \ref{Open-system squeezed-state framework}, the full density operator is
\begin{equation}
\rho = |O(\eta)\rangle\langle O(\eta)|.
\end{equation}
Tracing over the $-\vec{k}$ mode yields the reduced density matrix for mode $\vec{k}$,
\begin{equation}
\rho_k=\mathrm{Tr}_{-\vec k}\,\rho
=
\sum_{n=0}^{\infty} P_n\,|n_{\vec k}\rangle\langle n_{\vec k}|,
\label{rdm of otmss}
\end{equation}
where the eigenvalues are given by
\begin{equation}
P_n=
\frac{(1+u_2\tanh r_k)^2-|1-u_1^2|\tanh^2 r_k}{(1+u_2\tanh r_k)^2}
\,|1-u_1^2|^n
\frac{\tanh^{2n}r_k}{(1+u_2\tanh r_k)^{2n}}.
\end{equation}
Thus, the reduced density matrix is diagonal in the occupation-number basis and fully characterized by the spectrum $P_n$ whose details can be found in App. \ref{The calculation of RDM for OTMS}. This structure parallels the standard two-mode squeezed vacuum, but is modified by the open-system coefficients $u_1$ and $u_2$. For simplifying the notation, we define one parameter as follows, 
\begin{equation}
\alpha_k \equiv \frac{|1-u_1^2|\tanh^2 r_k}{(1+u_2\tanh r_k)^2},
\end{equation}
in terms of which
\begin{equation}
P_n=(1-\alpha_k)\alpha_k^n,
\qquad 0\le \alpha_k<1.
\end{equation}
The normalization condition $\sum_{n=0}^\infty P_n=1$ is then manifest.
In this form, all quantum-information measures considered below can be written compactly as functions of $\alpha_k$, or equivalently as functions of the squeezing parameter $r_k$ together with the open-system coefficients.

\subsection{Purity}

We first consider the purity of the reduced state,
\begin{equation}
\mu_k = \mathrm{Tr}(\rho_k^2).
\end{equation}
Using the diagonal form of $\rho_k$, one finds
\begin{equation}
\mu_k = \sum_{n=0}^{\infty} P_n^2
      = (1-\alpha_k)^2\sum_{n=0}^{\infty}\alpha_k^{2n}
      = \frac{1-\alpha_k}{1+\alpha_k}.
\end{equation}
Substituting the explicit form of $\alpha_k$ gives
\begin{equation}
\mu_k
=
\frac{
(1+u_2\tanh r_k)^2-|1-u_1^2|\tanh^2 r_k
}{
(1+u_2\tanh r_k)^2+|1-u_1^2|\tanh^2 r_k
}.
\label{purity}
\end{equation}
where the detail is found in App. \ref{Purity,von Neumann Entropy, Rényi Entropy, and Logarithmic Negativity}. The purity quantifies the degree of mixedness of the reduced state.
A pure reduced state corresponds to $\mu_k=1$, whereas $\mu_k<1$ signals mixedness.
In the present two-mode setting, this mixedness is induced by the correlations between the two modes.
Therefore, for a globally pure bipartite state, a decrease in purity indicates stronger entanglement between the subsystems.
More generally, purity provides a useful diagnostic of coherence loss and subsystem mixing, much as in the Gaussian formalism where purity is related to the covariance determinant via $\mu=\mathrm{Tr}(\rho^2)=1/\sqrt{\det V}$. In the language of decoherence, the purity is especially useful because it directly tracks the loss of reduced-state coherence after inaccessible degrees of freedom are traced out. When the off-diagonal components of the reduced density matrix are suppressed, the subsystem evolves from a nearly pure quantum state toward an effectively mixed state, and this transition is reflected by a decrease of $\mu_k$. Therefore, in the present open-system framework, a suppression of purity can be interpreted as a quantitative signature of decoherence in the observable sector \cite{Zurek:2003zz,Schlosshauer:2003zy}.

\subsection{von Neumann entropy}

The von Neumann entropy of the reduced state is defined by
\begin{equation}
S(\rho_k)=-\mathrm{Tr}(\rho_k\ln\rho_k)
        =-\sum_{n=0}^{\infty} P_n\ln P_n.
\end{equation}
For the geometric spectrum above, this becomes
\begin{equation}
S(\rho_k)
=
-\sum_{n=0}^{\infty}(1-\alpha_k)\alpha_k^n
\ln\bigl[(1-\alpha_k)\alpha_k^n\bigr].
\end{equation}
Evaluating the sums yields
\begin{equation}
S(\rho_k)
=
-\ln(1-\alpha_k)-\frac{\alpha_k}{1-\alpha_k}\ln \alpha_k.
\end{equation}
Equivalently, restoring the original variables,
\begin{equation}
S(\rho_k)
=
-\ln\!\left(
1-\frac{|1-u_1^2|\tanh^2 r_k}{(1+u_2\tanh r_k)^2}
\right)
-
\frac{
\frac{|1-u_1^2|\tanh^2 r_k}{(1+u_2\tanh r_k)^2}
}{
1-\frac{|1-u_1^2|\tanh^2 r_k}{(1+u_2\tanh r_k)^2}
}
\ln\!\left(
\frac{|1-u_1^2|\tanh^2 r_k}{(1+u_2\tanh r_k)^2}
\right), 
\label{van neumann entropy}
\end{equation}
whose detail is found in App. \ref{Purity,von Neumann Entropy, Rényi Entropy, and Logarithmic Negativity}.
This entropy measures the uncertainty or mixedness of the reduced mode.
When the total bipartite state is pure, the von Neumann entropy of the reduced density matrix is equal to the entanglement entropy.
However, this identification is no longer valid once the total state becomes mixed; in that case, the von Neumann entropy measures only the entropy of the subsystem and is not, by itself, a reliable entanglement measure.
This distinction will be important when interpreting the physical role of the open-system parameters. From the viewpoint of decoherence, the von Neumann entropy measures the growth of uncertainty in the reduced state caused by tracing out the environment. For a globally pure bipartite state, this entropy coincides with the entanglement entropy, but for an effectively open or mixed global state it also receives contributions from classical statistical mixing. Hence, an increase in $S(\rho_k)$ should be interpreted as evidence for entropy production and information loss in the reduced description, rather than automatically as a one-to-one measure of entanglement \cite{von2013mathematische}.

\subsection{R\'enyi entropy}

A useful generalization of the von Neumann entropy is the R\'enyi entropy of order $\mu$,
\begin{equation}
S_\mu(r_k)=\frac{1}{1-\mu}\ln\left(\sum_{n=0}^{\infty}P_n^\mu\right),
\qquad \mu\ge 0.
\end{equation}
Using the spectrum above, one obtains
\begin{equation}
S_\mu(r_k)
=
\frac{1}{1-\mu}
\ln\left[
\sum_{n=0}^{\infty}
\left(
\frac{(1+u_2\tanh r_k)^2-|1-u_1^2|\tanh^2 r_k}{(1+u_2\tanh r_k)^2}
\,|1-u_1^2|^n
\frac{\tanh^{2n}r_k}{(1+u_2\tanh r_k)^{2n}}
\right)^\mu
\right].
\end{equation}
After summing the geometric series, this reduces to
\begin{equation}
S_\mu(r_k)
=
\frac{1}{1-\mu}
\left[
\mu\ln(1-\alpha_k)-\ln(1-\alpha_k^\mu)
\right],
\end{equation}
or explicitly,
\begin{equation}
S_\mu(r_k)
=
\frac{1}{1-\mu}
\left[
\mu\ln\left(
1-\frac{|1-u_1^2|\tanh^2 r_k}{(1+u_2\tanh r_k)^2}
\right)
-\ln\left(
1-\frac{|1-u_1^2|^\mu\tanh^{2\mu}r_k}{(1+u_2\tanh r_k)^{2\mu}}
\right)
\right].
\label{Smu}
\end{equation}

Two especially useful special cases are
\begin{equation}
S_2(r_k)
=
-2\ln(1-\alpha_k)+\ln(1-\alpha_k^2)
=
\ln\left(\frac{1+\alpha_k}{1-\alpha_k}\right),
\end{equation}
and
\begin{equation}
S_{1/2}(r_k)
=\ln\left(\frac{1+\sqrt{\alpha_k}}{1-\sqrt{\alpha_k}}\right).
\end{equation}
In terms of $u_1$, $u_2$, and $r_k$, these become
\begin{equation}
\begin{split}
S_2(r_k)
&=
-2\ln\left(
1-\frac{|1-u_1^2|\tanh^2 r_k}{(1+u_2\tanh r_k)^2}
\right)
+\ln\left(
1-\frac{|1-u_1^2|^2\tanh^4 r_k}{(1+u_2\tanh r_k)^4}
\right)\\&=\ln{(\frac{1+\frac{|1-u_1^2|\tanh^2 r_k}{(1+u_2\tanh r_k)^2}}{1-\frac{|1-u_1^2|\tanh^2 r_k}{(1+u_2\tanh r_k)^2}})},
\label{S20}
\end{split}
\end{equation}
and
\begin{equation}
\begin{split}
S_{1/2}(r_k)
&=
\ln\left(
1-\frac{|1-u_1^2|\tanh r_k^2}{(1+u_2\tanh r_k)^2}
\right)
-2\ln\left(
1-\frac{|1-u_1^2|^{\frac{1}{2}}\tanh r_k}{1+u_2\tanh r_k}
\right)\\&=\ln{(\frac{1+\frac{|1-u_1^2|^{\frac{1}{2}}\tanh r_k}{1+u_2\tanh r_k}}{1-\frac{|1-u_1^2|^{\frac{1}{2}}\tanh r_k}{1+u_2\tanh r_k}})}.
\label{S120}
\end{split}
\end{equation}
The detail is found in App. \ref{Purity,von Neumann Entropy, Rényi Entropy, and Logarithmic Negativity}.
The R\'enyi entropy reduces to the von Neumann entropy in the limit $\mu\to1$.
Moreover, by analogy with the standard Gaussian case, the second- and half-order R\'enyi entropies provide lower and upper bounds on the von Neumann entropy,
\begin{equation}
S_2 \le S_1 \le S_{1/2},
\label{relation between two entropies}
\end{equation}
with $S_1$ denoting the von Neumann entropy.
This is particularly useful in the large-squeezing regime, where the direct numerical evaluation of $S_1$ may become unstable, while $S_2$ and $S_{1/2}$ remain numerically robust. The R\'enyi entropies are also useful from the perspective of decoherence because they probe the spectrum of the reduced density matrix with different sensitivities. In particular, the second-order R\'enyi entropy is directly related to the purity through $S_2=-\ln \mu_k$, and therefore provides an alternative measure of reduced-state mixedness. More generally, the family of R\'enyi entropies captures how the eigenvalue distribution of the reduced density matrix broadens during decoherence, making them valuable complementary diagnostics of entropy production in cosmological open systems \cite{renyi1961measures}.

\subsection{Logarithmic negativity}

To quantify the bipartite entanglement between the modes $\vec{k}$ and $-\vec{k}$, we consider the logarithmic negativity.
This quantity is based on the positivity of partial transposition (PPT) criterion and is defined by
\begin{equation}
E_{\mathcal N}=\ln\|\rho^{\,T_{-\vec k}}\|_1,
\end{equation}
where $\rho^{\,T_{-\vec k}}$ denotes the partial transpose of the full density matrix with respect to the $-\vec{k}$ mode, and $\|\cdot\|_1$ is the trace norm.
Equivalently,
\begin{equation}
E_{\mathcal N}
=
\ln\left(\sum_i |\lambda_i^{\mathrm{pt}}|\right),
\end{equation}
where $\lambda_i^{\mathrm{pt}}$ are the eigenvalues of the partially transposed density matrix.

For the present open two-mode squeezed-state construction, the logarithmic negativity is determined by the same squeezing data that appear in the reduced density matrix.
In the weak-dissipation limit, it reduces to the standard two-mode squeezed result and is therefore a monotonic function of the squeezing parameter $r_k$.
This is consistent with the Gaussian-state result that logarithmic negativity directly diagnoses the entanglement generated between the $(\vec{k},-\vec{k})$ modes, unlike the von Neumann entropy, whose interpretation depends on whether the global state is pure or mixed.

In the present framework, the explicit expression can be derived ,
\begin{equation}
E_{\mathcal N}=\log{\frac{(1+ u_2\tanh{r_k}+\tanh{r_k}|1-u_1^2|^{\frac{1}{2}})}{(1+ u_2\tanh{r_k}-\tanh{r_k}|1-u_1^2|^{\frac{1}{2}})}},
\label{negativity}
\end{equation}
and thus inherits the effects of nontrivial sound speed implicitly through the Schr\"odinger evolution of the squeezing parameter and the detail is found in App. \ref{Purity,von Neumann Entropy, Rényi Entropy, and Logarithmic Negativity}.
As in the cases of purity and entropy, the sound-speed dependence enters dynamically through $r_k$, rather than through a new algebraic structure of the entanglement measure itself. Unlike purity and entropic quantities, which mainly diagnose mixedness and information loss in the reduced sector, the logarithmic negativity is designed to isolate genuinely quantum bipartite correlations. This distinction is important in decohering systems: decoherence can increase mixedness and entropy while simultaneously reducing the distillable quantum correlation between subsystems. Therefore, the logarithmic negativity provides a complementary probe of how much nonclassical entanglement survives the decohering dynamics of the $(\vec{k},-\vec{k})$ pair \cite{Plenio:2005cwa}.


\subsection{Interpretation and diagnostic role}

The four quantities introduced above should be viewed as complementary diagnostics rather
than redundant observables. Purity measures the degree of reduced-state mixedness and
therefore provides the most immediate indicator of coherence loss after tracing out one
sector. The von Neumann entropy quantifies the basis-independent uncertainty of the
reduced state and, for a globally pure bipartite system, coincides with the entanglement
entropy. The R\'enyi entropies generalize this information to a one-parameter family and are
especially useful because they probe the spectrum of $\rho_k$ with different sensitivities; in
particular, the second-order R\'enyi entropy is directly related to the purity through
$S_2=-\ln\mu_k$. Finally, the logarithmic negativity plays a distinct role: because it is based
on the partial-transpose criterion, it directly diagnoses bipartite entanglement even in
situations where the total state is effectively mixed \cite{Zurek:2003zz,Schlosshauer:2003zy,Vidal:2002zz,Peres:1996dw}.

From the viewpoint of decoherence, these quantities separate several notions that are often
conflated. A decrease in purity and an increase in entropic measures indicate that the
reduced description has become more mixed and more classicalized, in the sense that
phase information has been transferred to inaccessible degrees of freedom. However, this
does not by itself determine how much genuine quantum entanglement remains between the
$\vec{k}$ and $-\vec{k}$ modes. For this reason, the logarithmic negativity is particularly
important: it distinguishes residual bipartite quantum correlation from mixedness alone.
This distinction has long been central in discussions of the quantum-to-classical transition of
primordial perturbations, where decoherence selects an effectively classical reduced
description while the underlying state still originates from quantum squeezing
\cite{Polarski:1995jg,Kiefer:1998qe,Lombardo:2005iz,Brandenberger:1990bx}.

In the present Gaussian two-mode framework, all these quantities are ultimately determined
by the same reduced-state spectrum, or equivalently by the squeezing data and the open-system
coefficients. Their main value is therefore diagnostic: they reveal different aspects of the same
underlying dynamics. The purity tracks the loss of reduced-state coherence, the von Neumann
and R\'enyi entropies characterize the accompanying entropy production, and the logarithmic
negativity isolates the genuinely entangled part of the $(\vec{k},-\vec{k})$ correlation. Since the
nontrivial sound speed modifies the Schr\"odinger evolution of the squeezing variables, its
physical imprint is inherited by all these measures dynamically rather than algebraically.
This observation will guide our interpretation of the numerical results in the next section.

\section{Numerical results in inflation}
\label{Numerical results in the inflationary regime}
Following the discussion in Sec. \ref{SSR in inflation}, we utilize the bounded variable $x=\tanh r_k$ in place of $r_k$ to investigate the system's evolution. 
This substitution maps the squeezing parameter $r_k$ onto a bounded interval. In practice, we restrict our numerical analysis to the interval $-1 \le y = \log_{10} a \le 0$, which is sufficient to capture the essential dynamics during the inflationary era.

\subsection{Inflationary evolution of \texorpdfstring{$x=\tanh r_k$}{x=tanh rk} and \texorpdfstring{$\phi_k$}{phik}}

We begin by examining the two dynamical variables that govern the reduced-state observables: the bounded squeezing variable $x = \tanh r_k$ and the squeezing phase $\phi_k$. Substituting $x$ for the squeezing amplitude $r_k$ maps the otherwise unbounded growth of $r_k$ onto the finite interval $|x| < 1$.

\begin{figure}[h]
    \centering
    \includegraphics[width=.4\textwidth]{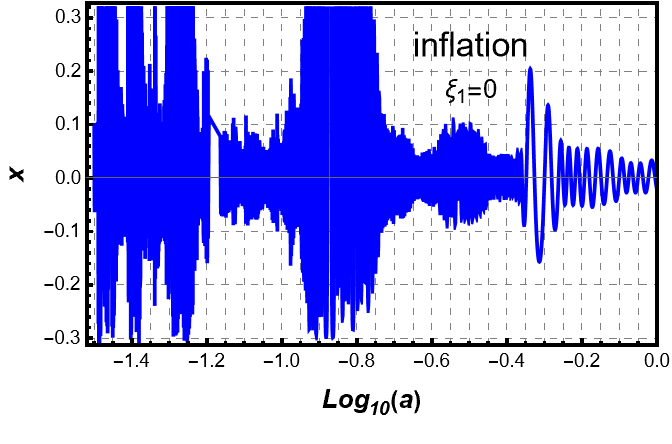}
    \qquad
    \includegraphics[width=.4\textwidth]{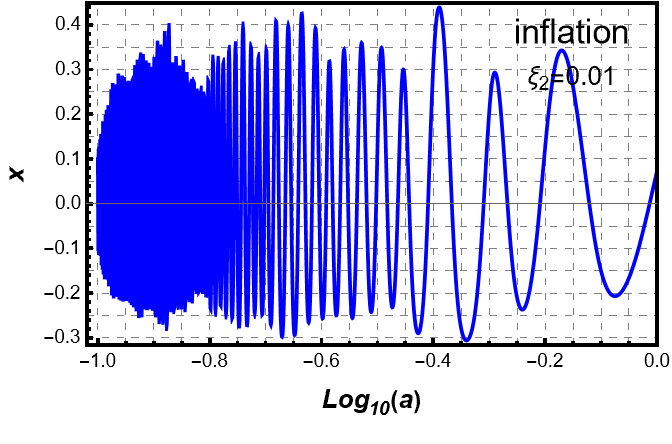}
    \qquad
    \includegraphics[width=.4\textwidth]{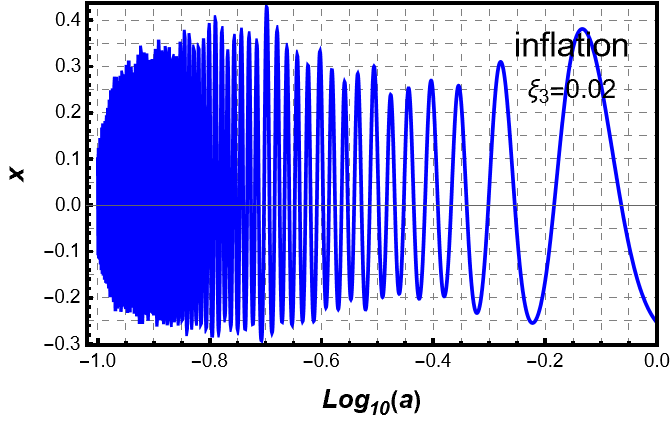}
     \qquad
    \includegraphics[width=.4\textwidth]{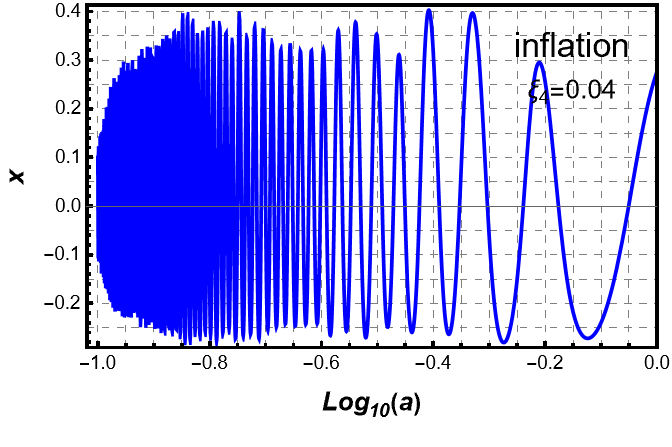}
    \caption{The numerical results of $r_{k}$ in terms of $\log_{10}a$ for inflation, where we have $k=1$ and $H_0=1$ for simplicity. }
    \label{fig: op x}
\end{figure}

\begin{figure}[h]
    \centering
    \includegraphics[width=.4\textwidth]{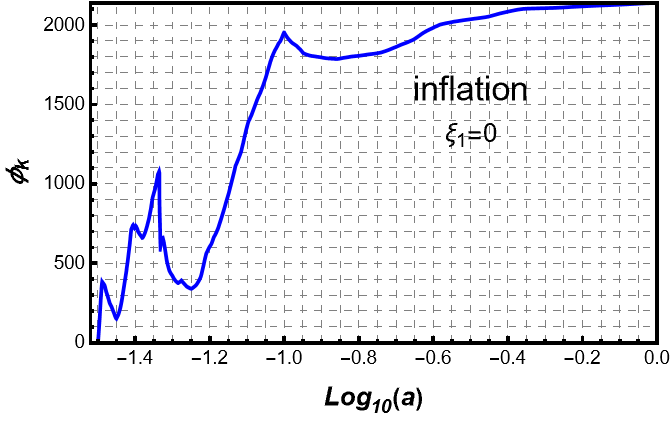}
    \qquad
    \includegraphics[width=.4\textwidth]{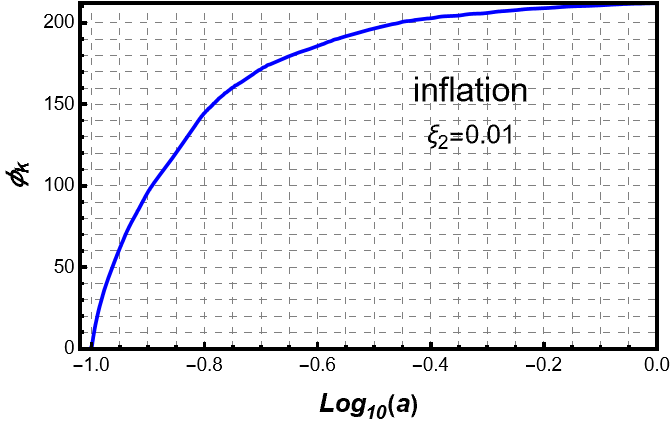}
    \qquad
    \includegraphics[width=.4\textwidth]{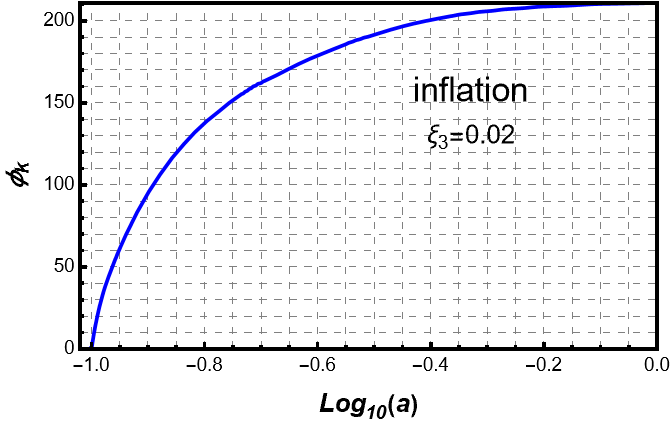}
     \qquad
    \includegraphics[width=.4\textwidth]{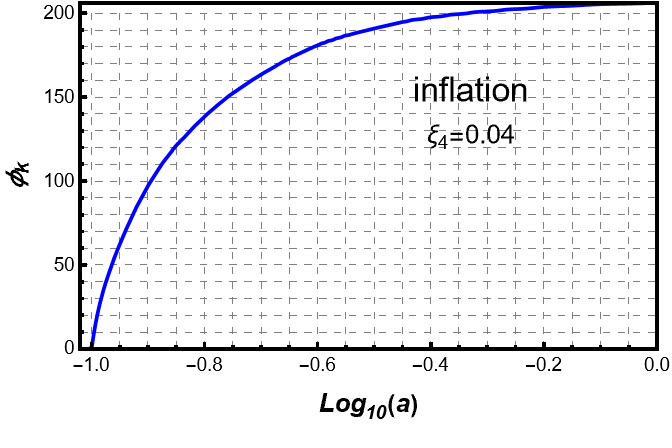}
    \caption{The numerical results of $\phi_{k}$ in terms of $\log_{10}a$ for inflation, where we have $k=1$ and $H_0=1$ for simplicity.}
    \label{fig: op phi_k}
\end{figure}
Within the interval $-1 \le y \le 0$, Figs. \ref{fig: op x} and \ref{fig: op phi_k} demonstrate that both $x$ and $\phi_k$ exhibit pronounced oscillatory behavior, which is strongly modulated by the sound-speed parameter $\xi$. 
Fig. \ref{fig: op x} shows that the regular oscillations seen in the standard case become increasingly perturbed as $\xi$ increases. This figure clearly indicates that a non-trivial sound speed significantly modifies the evolution of the squeezing amplitude $r_k$.A similar trend is observed for the squeezing phase $\phi_k$ in Fig. \ref{fig: op phi_k}. While the standard case displays irregular oscillatory behavior, the introduction of a non-trivial sound speed dramatically reduces the amplitude of $\phi_k$—by a factor of up to 200—relative to the standard scenario. Based on these two numerical results, we can now proceed with the information-theoretic diagnostics of the cosmological perturbations.

\subsection{Purity}

\begin{figure}
    \centering
    \includegraphics[width=.8\textwidth]{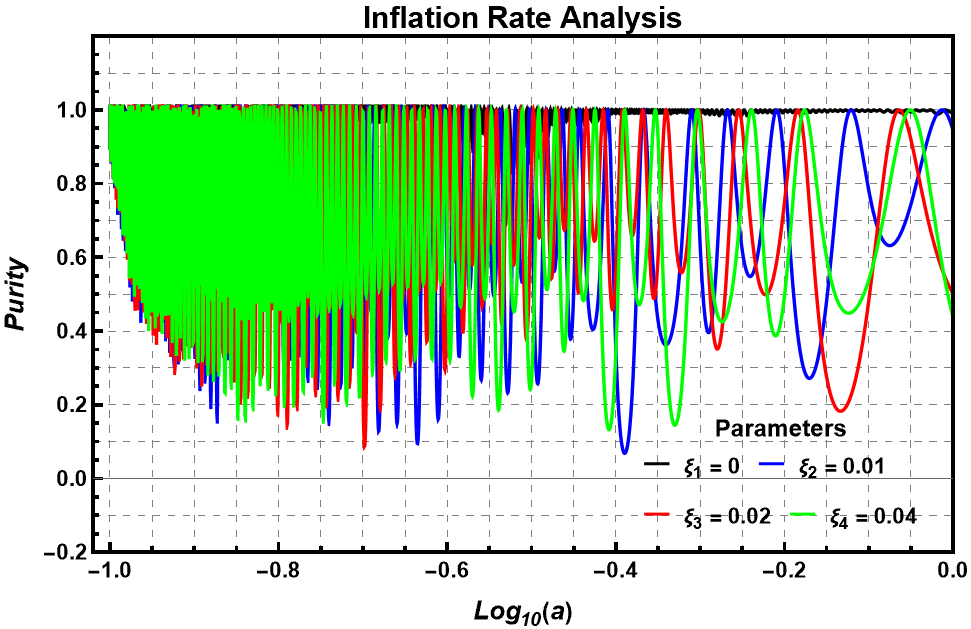}
    \qquad
    \caption{The numerical results of Purity in terms of $\log_{10}a$ for inflation, where we have set  $k=1$ and $H_0=1$. }
    \label{fig: op s purity}
\end{figure}

We next consider the purity of the RDM \eqref{rdm of otmss},
\begin{equation}
\mu_k=\mathrm{Tr}\,\rho_k^2,
\end{equation}
which measures the mixedness of the reduced state.
The Fig. \ref{fig: op s purity} shows that in the inflationary regime the purity exhibits strong oscillatory behavior below unity, indicating that the reduced mode is substantially mixed.
This is consistent with the physical interpretation of a correlated two-mode system, where tracing out one momentum sector produces a mixed reduced state even when the full state is generated from a squeezed construction.

As illustrated in Fig. \ref{fig: op s purity}, the purity in the standard case remains near unity, indicating that the RDM remains essentially pure. However, increasing the value of $\xi$ leads to significant departures from this standard behavior. Physically, this implies that a non-trivial sound speed enhances the effective subsystem mixedness during inflation. Since purity decreases as a state becomes more mixed, the suppression of $\mu_k$ indicates that sound-speed resonance (SSR) strengthens the correlations encoded within OTMSS \eqref{normalized otmss}.

\subsection{von Neumann entropy and R\'enyi entropies}

\begin{figure}
    \centering
    \includegraphics[width=.7\textwidth]{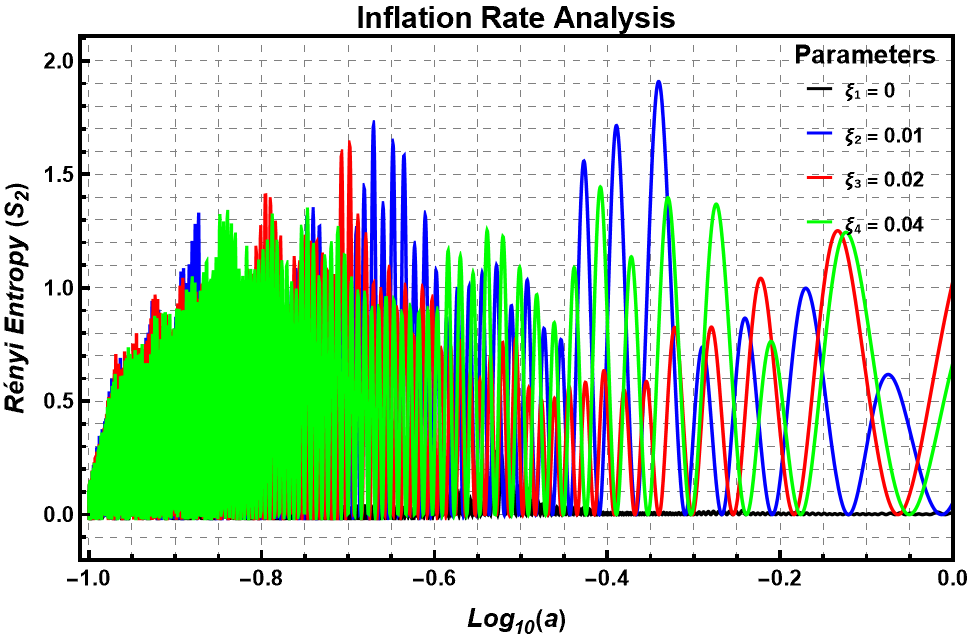}
    \qquad
     \includegraphics[width=.7\textwidth]{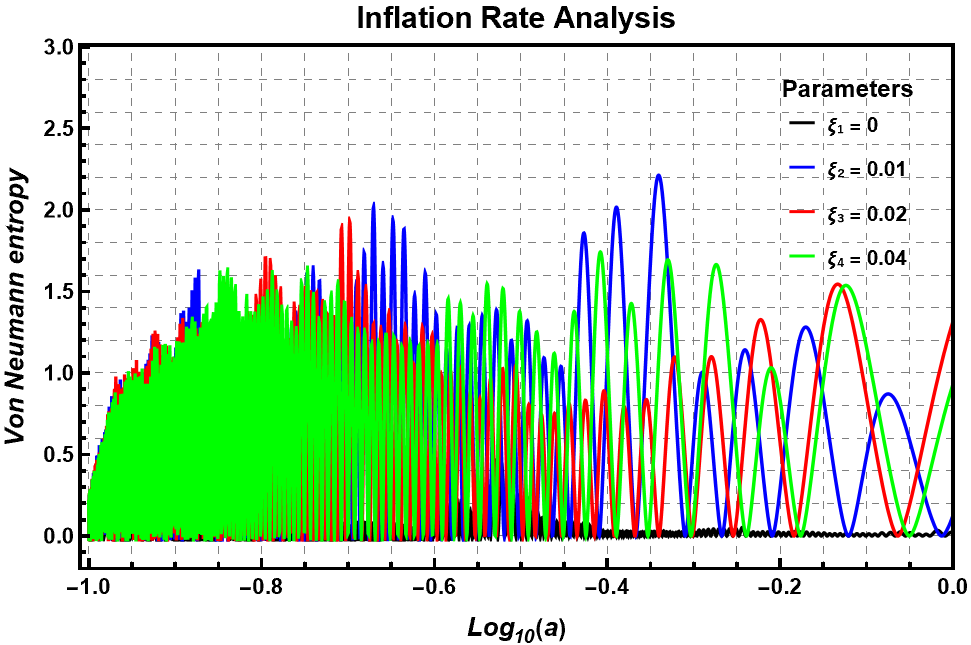}\\
    \qquad
     \includegraphics[width=.7\textwidth]{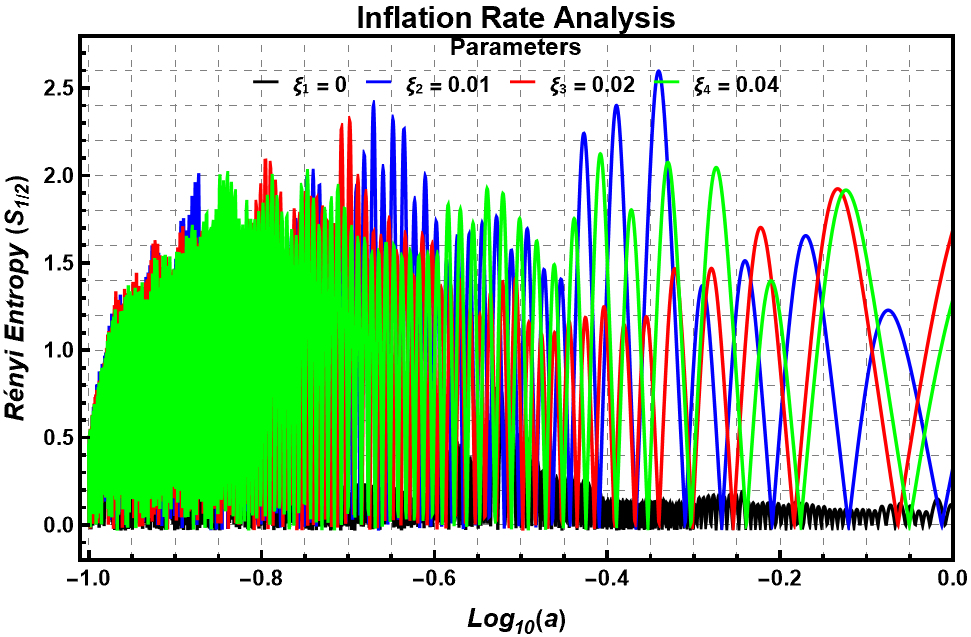}
    \qquad
    \caption{The numerical results of Rényi Entropy and Von Neumann entropy in terms of $\log_{10}a$ for inflation, where we have set $k=1$ and $H_0=1$. }
    \label{fig: op s Rényi Entropy and Von Neumann entropy}
\end{figure}

In this section, we present the numerical results for the von Neumann and Rényi entropies based on Eqs. \eqref{van neumann entropy}, \eqref{S20}, and \eqref{S120}, as illustrated in Fig. \ref{fig: op s Rényi Entropy and Von Neumann entropy}. The behavior of the von Neumann entropy is strictly complementary to that of the purity: as the reduced state becomes increasingly mixed, the entropy rises accordingly. As shown in the second panel of Fig. \ref{fig: op s Rényi Entropy and Von Neumann entropy}, the von Neumann entropy develops a pronounced modulation for non-zero values of $\xi$. In contrast, the standard case ($\xi=0$) remains comparatively small throughout most of the interval. These results demonstrate that a non-trivial sound speed significantly enhances entropy production within the reduced sector during the inflationary epoch.

These conclusions are further corroborated by the Rényi entropies. Both the second-order ($q=2$) and half-order ($q=1/2$) Rényi entropies exhibit a clear sensitivity to the sound-speed parameter; larger values of the non-trivial sound speed lead to larger entropy values and more pronounced variations. This finding is significant for two primary reasons. First, it demonstrates that the sound-speed imprint is a robust physical feature rather than an artifact of a specific entropy definition. Second, the Rényi entropies are numerically advantageous, as they offer greater stability than the direct evaluation of the von Neumann entropy, particularly in regimes with pronounced modulation. Consequently, they provide a reliable characterization of the entropy growth associated with the inflationary squeezing process. A key feature of the present results is the close alignment between the trends of the von Neumann and Rényi entropies. This is to be expected, as both are determined by the same squeezing parameters within the RDM derived from the normalized wave function. Rather than representing independent dynamical degrees of freedom, these measures offer complementary information-theoretic perspectives on the same underlying inflationary evolution. Furthermore, the relative magnitudes of the von Neumann and Rényi entropies remain consistent with the analytical relation defined in \eqref{relation between two entropies}.

\subsection{Logarithmic negativity}
\label{Logarithmic negativity}

The logarithmic negativity \eqref{negativity} serves as the most direct entanglement-oriented observable in our analysis. While the interpretation of the von Neumann entropy as an entanglement measure is contingent upon whether the global state remains pure or becomes mixed, the logarithmic negativity provides a robust diagnosis of bipartite entanglement specifically between the $\vec{k}$ and $-\vec{k}$ modes.

\begin{figure}
    \centering
    \includegraphics[width=.8\textwidth]{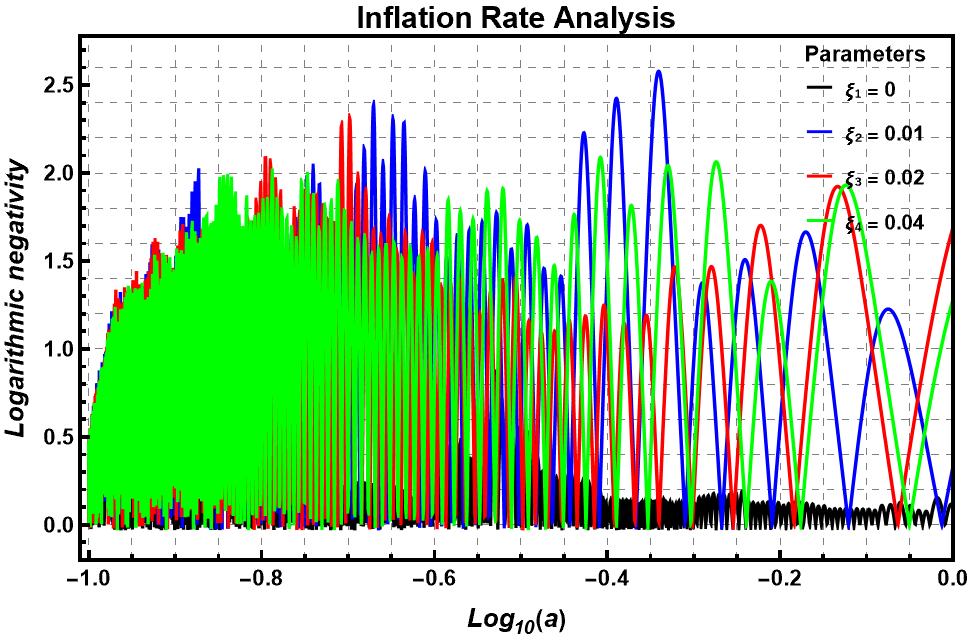}
    \qquad
    \caption{The numerical results of Logarithmic negativity in terms of $\log_{10}a$ for inflation, where we have set $k=1$ and $H_0=1$. }
    \label{fig: op Logarithmic negativity}
\end{figure}

Numerically, the logarithmic negativity exhibits pronounced modulation throughout the inflationary interval, as illustrated in Fig. \ref{fig: op Logarithmic negativity}. This behavior represents a significant deviation from the standard case, for which the logarithmic negativity remains comparatively small. As the parameter $\xi$ increases, both the logarithmic negativity and the degree of its modulation increase accordingly. This demonstrates that SSR affects not only the subsystem mixedness and entropy production but also the fundamental entanglement structure of the two-mode state. Consequently, among the observables considered in this study, the logarithmic negativity provides the most distinct entanglement-based signature of non-trivial sound speed during inflation.

It is also noteworthy that the results for logarithmic negativity are qualitatively consistent with those for entropy: cases exhibiting more pronounced non-trivial sound-speed effects tend to display larger deviations from standard behavior in both sectors. This consistency reinforces the interpretation that these features originate from the modified squeezing dynamics induced by the sound-speed-dependent Hamiltonian.

\subsection{Overall interpretation}

The numerical results collectively support a consistent inflationary paradigm. Our analysis uses the bounded variable $x=\tanh r_k$ within the inflationary interval $-1 \le y \le 0$. In this regime, the impact of non-trivial sound speed is unambiguous: it modifies the oscillatory evolution of the squeezing variables and leaves distinct imprints on all reduced-state observables. To understand the physical implications, we examine four key quantities: purity, von Neumann entropy, Rényi entropy, and logarithmic negativity. The observed suppression of purity indicates increased mixedness within the two-mode system. Conversely, the von Neumann entropy, Rényi entropy, and logarithmic negativity all increase. These quantities characterize the system's decoherence; specifically, higher entropic values represent a more significant departure from quantum purity. Taken together, these findings suggest a clear physical picture: a non-trivial sound speed postpones the onset of classicality by modulating the decoherence process.

\section{Conclusions and outlook}
 \label{Conclusions}

In this work, we have investigated the quantum-information diagnostics of cosmological perturbations with a non-trivial sound speed within an open-system squeezed-state framework. By deriving the RDM from the normalized wave function, we analyzed key observables including purity, von Neumann entropy, Rényi entropies, and logarithmic negativity. Our central finding is that the impact of a non-trivial sound speed enters these observables dynamically rather than algebraically. The SSR reshapes the Schrödinger evolution of the state, thereby altering the inflationary trajectory of the squeezing variables. Consequently, these dynamical modifications are inherited by the reduced-state diagnostics, leaving identifiable quantum-information signatures—such as suppressed purity and enhanced entanglement—during the inflationary era. Our main findings are summarized as follows: 

 $(1)$ \textbf{Dynamical Sensitivity of Squeezing Variables ($r_k$ and $\phi_k$)}: The numerical results demonstrate that a non-trivial sound speed significantly modifies the oscillatory evolution of the squeezing parameters. The squeezing phase ($\phi_k$) is exceptionally sensitive to the SSR; the introduction of SSR dramatically reduces the amplitude of the phase by a factor of up to 200 relative to the standard canonical scenario ($\xi=0$).

 $(2)$ \textbf{Suppression of Purity and Enhanced Mixedness}: The purity of the RDM is notably suppressed by the non-trivial sound speed, exhibiting strong oscillatory behavior well below unity. Physically, this suppression indicates that the SSR strengthens the correlations between the observed and traced-out momentum sectors, leading to enhanced effective subsystem mixedness and a more significant departure from quantum purity. 

 $(3)$ \textbf{Enhancement of Entropy Production}: Both the von Neumann entropy and the Rényi entropies ($S_2$ and $S_{1/2}$) are significantly enhanced and exhibit pronounced oscillatory modulations for non-zero values of the sound-speed parameter ($\xi$). This confirms that a non-trivial sound speed accelerates entropy production within the reduced sector. Furthermore, the Rényi entropies prove to be numerically advantageous, offering a more stable characterization of entropy growth in strongly oscillatory regimes than the direct evaluation of the von Neumann entropy.

 $(4)$ \textbf{Modulation of Bipartite Entanglement (Logarithmic Negativity)}: As the most direct diagnostic of genuine bipartite quantum entanglement between the $\mathbf{k}$ and $-\mathbf{k}$ modes, the logarithmic negativity exhibits highly oscillatory patterns with amplitudes that grow alongside the parameter $\xi$. This demonstrates that the sound-speed resonance does not merely increase statistical mixedness; it fundamentally reshapes the survival and structure of nonclassical entanglement during the decohering dynamics of inflation.

 \textbf{Future Research Outlook:} To transcend the current analytical limitations, our future research will advance in several promising directions. First, we aim to quantitatively analyze the decoherence effects of SSR by employing the RDM covariance matrix method as a function of the e-folding number. Second, we plan to implement advanced lattice methods to simulate the quantum-information evolution throughout the RD and MD epochs. Furthermore, our formalism can be naturally extended to incorporate $f(R)$ gravity \cite{Liu:2018hno,Liu:2018htf} and multi-field inflationary models \cite{Liu:2019xhn,Liu:2020zzv,Liu:2020zlr,Liu:2021rgq,Zhang:2022bde}. Finally, because the expansion of the universe can be formally treated as a non-inertial frame, we intend to generalize our framework to explore quantum-information dynamics in broader non-inertial systems \cite{Liu:2025hcx,Tang:2025mtc,Jiang:2025ktt,Wu:2025wvb,Wu:2024yop}.

\acknowledgments

LH, BC and SC are funded by NSFC grant NO. 12665009 and Hunan Provincial Department of Education Project NO. 25B0480. HQ is funded by NSFC grant NO. 12675057. PZ is funded by NSFC grant NO. 1260051567, the Key Project of Sichuan Science and Technology Education Joint Fund (No. 25LHJJ0097), and the Sichuan Natural Science Foundation (Grant No. 2026NSFSC0746).

\appendix

\section{The normalization of OTMSS}
\label{normalization of otmss}
In this appendix, we provide the detailed derivation for the normalization of the OTMSS. We begin by introducing the normalization factor $N$ as follows:
\begin{equation}
    \begin{split}
        N\ket{O(\eta)}=N\frac{ \rm sech r_k(\eta)}{1+u_2\tanh r_k(\eta)}\sum_{n=0}^{\infty}|1-u_1^2|^{\frac{n}{2}}\frac{(-\exp(2i\phi_k(\eta))\tanh r_k(\eta))^{n}}{(1+u_2\tanh r_k( \eta))^{n}}|n,n\rangle_{-\vec{k},\vec k}.
    \end{split}
    \label{N in OTMSS}
\end{equation}
Then, according to the normalization condition $\langle O(\eta) | N^2 | O(\eta) \rangle = 1$, we have: 
\begin{equation}
    \begin{split}
        & N^2\frac{ \rm sech^2r_k(\eta)}{(1+u_2\tanh r_k(\eta))^2}\sum_{n=0}^{\infty}|1-u_1^2|^{n}\frac{(\tanh r_k(\eta))^{2n}}{(1+u_2\tanh r_k( \eta))^{2n}}=1
    \end{split}
\end{equation}
Thus, we could derive the formula of $N$ as follows,
\begin{equation}
    N=\frac{\sqrt{(1+u_2\tanh{r_k})^2-|1-u_1^2|\tanh^2{r_k}}}{\mbox{sechr}_k}. 
\end{equation}
By combining this with Eq. \eqref{N in OTMSS}, we can readily derive the normalized OTMSS as given in Eq. \eqref{normalized otmss}.

\section{The evolution of $r_k$ and $\phi_k$}
\label{rk and phhi}
In this appendix, we will calculate the evolution equation for $r_k$ and $\phi_k$ according to the Schr\"odinger equation as follows,
\begin{equation}
\hat H\,|O(\eta)\rangle = i\,\partial_\eta |O(\eta)\rangle,
\label{schrodinger equation}
\end{equation}
where $\hat{H}$ is the Hamiltonian operator defined in Eq. \eqref{eq:Hk_final}. In the subsequent analysis, we decompose the state into its ground-state and excited-state components. Focusing first on the ground state, we apply the Hamiltonian to the vacuum state to derive the corresponding ground-state expression as follows:
\begin{equation}    
\begin{split}        
=&(\frac{\sqrt{(1+u_2\tanh{r_k})^2-|1-u_1^2|\tanh^2{r_k}}}{1+u_2\tanh(r_k)})[(\frac{a^2V_{\phi\phi}}{2k}+\frac{k}{2}+\frac{kc_s^2}{2})-(\frac{a^2V_{\phi\phi}}{2k}-\frac{z'}{z}i-\frac{k}{2}+\frac{kc_s^2}{2})\\&|1-u_1^2|^{\frac{1}{2}}e^{2i\phi_k}(\frac{\tanh(r_k)}{1+u_2\tanh(r_k)})] \ket {0,0}_{\vec{k},-\vec{k}}    
\end{split}
\label{ground state of hamiltonian}
\end{equation}
Furthermore, we can derive the excited-state expression by isolating the terms that are independent of $n$, 
\begin{equation}
    \begin{split}
        =&(\frac{\sqrt{(1+u_2\tanh{r_k})^2-|1-u_1^2|\tanh^2{r_k}}}{1+u_2\tanh(r_k)})[(\frac{a^2V_{\phi\phi}}{2k}+\frac{k}{2}+\frac{kc_s^2}{2})-(\frac{a^2V_{\phi\phi}}{2k}-\frac{z'}{z}i-\frac{k}{2}+\frac{kc_s^2}{2})\\&|1-u_1^2|^{\frac{1}{2}}e^{2i\phi_k}(\frac{\tanh(r_k)}{1+u_2\tanh(r_k)})] \sum_{n=1}^\infty (-1)^n |1-u_1^2|^\frac{n}{2}e^{2in\phi_k}\\&(\frac{\tanh(r_k)}{1+u_2\tanh(r_k)})^n\ket {n,n}_{\vec{k},-\vec{k}}.
    \end{split}
    \label{excited state without n dependent term}
\end{equation}
Regarding the excited-state component containing the $n$-dependent terms, the resulting formula can be straightforwardly derived as follows:
\begin{equation}    
\begin{split}        
\\=&(\frac{\sqrt{(1+u_2\tanh{r_k})^2-|1-u_1^2|\tanh^2{r_k}}}{1+u_2\tanh(r_k)})[2i\phi'_k+\frac{r'_k\rm{sech}^2r_k}{\tanh{r_k}}+\frac{|1-u_1^2|'}{2|1-u_1^2|}\\&-\frac{u'_2\tanh{r_k}+u2r_k'\rm{sech}^2r_k}{1+u_2\tanh{r_k}}]\sum_{n=1}^\infty (-1)^n |1-u_1^2|^\frac{n}{2}e^{2in\phi_k}(\frac{\tanh(r_k)}{1+u_2\tanh(r_k)})^nn\ket {n,n}_{\vec{k},-\vec{k}}.    \end{split}
\label{excited term with n dependent term}
\end{equation}
Subsequently, we apply the operator $i\partial_\eta$ to the normalized OTMSS \eqref{normalized otmss}, following a similar logic to the previous derivation. We begin by considering the ground-state component:
\begin{equation}    
\begin{split}        
=i\frac{(r_k'+u_2')|1-u_1^2|\tanh^3{r_k}-(1+u_2\tanh{r_k})\frac{1}{2}|1-u_1^2|'\tanh^2{r_k}-r_k'|1-u_1^2|\tanh{r_k}}{(1+u_2\tanh{r_k})^2\sqrt{(1+u_2\tanh{r_k})^2-|1-u_1^2|\tanh^2{r_k}}}\ket {0,0}_{\vec{k},-\vec{k}}]    \end{split}
\label{ground state for right side}
\end{equation}
where $u_2$ and $\sqrt{|1-u_1^2|}$ are both functions of conformal time in our framework. For the excited states, the component independent of the $n$-th index is obtained as follows:
\begin{equation}    
\begin{split}        
=&i\frac{(r_k'+u_2')|1-u_1^2|\tanh^3{r_k}-(1+u_2\tanh{r_k})\frac{1}{2}|1-u_1^2|'\tanh^2{r_k}-r_k'|1-u_1^2|\tanh{r_k}}{(1+u_2\tanh{r_k})^2\sqrt{(1+u_2\tanh{r_k})^2-|1-u_1^2|\tanh^2{r_k}}}\\&\sum^{\infty}_{n=1}(-1)^n|1-u_1^2|^{\frac{n}{2}}e^{2in\phi_k}(\frac{\tanh(r_k)}{1+u_2\tanh(r_k)})^n\ket {n,n}_{\vec{k},-\vec{k}}].    \end{split}
\label{without nth term on the right}
\end{equation}
As for the piece with $n-th$ dependent term, it is derived as follows,
\begin{equation}    
\begin{split}        
=&(\frac{\sqrt{(1+u_2\tanh{r_k})^2-|1-u_1^2|\tanh^2{r_k}}}{1+u_2\tanh(r_k)})[(\frac{a^2V_{\phi\phi}}{2k}+\frac{k}{2}+\frac{kc_s^2}{2})-(\frac{a^2V_{\phi\phi}}{2k}-\frac{z'}{z}i-\frac{k}{2}+\frac{kC_s^2}{2})\\&|1-u_1^2|^{\frac{1}{2}}e^{2i\phi_k}(\frac{\tanh(r_k)}{1+u_2\tanh(r_k)})-(\frac{a^2V_{\phi\phi}}{2k}-\frac{z'}{z}i-\frac{k}{2}+\frac{kc_s^2}{2})|1-u_1^2|^{-\frac{1}{2}}e^{-2i\phi_k}\\&(\frac{\tanh(r_k)}{1+u_2\tanh(r_k)})^{-1}]\sum_{n=1}^\infty (-1)^n |1-u_1^2|^\frac{n}{2}e^{2in\phi_k}\\&(\frac{\tanh(r_k)}{1+u_2\tanh(r_k)})^nn\ket {n,n}_{\vec{k},-\vec{k}}.    \end{split}
\label{with nth term on the right}
\end{equation}
Finally, by equating the terms on both sides of the Schrödinger equation, we obtain the evolution equations for $r_k$ and $\phi_k$ as follows:
\begin{equation}    
\begin{split}        
r'_k&=\frac{|1-u_1^2|^{\frac{1}{2}}[(1+u_2\tanh r_k)^2-|1-u_1^2|\tanh^2{r_k}][\frac{z'}{z}\cos{2\phi_k}-(\frac{a^2V_{\phi\phi}}{2k}-\frac{k}{2}+\frac{kc_s^2}{2})\sin{2\phi_k}]}{|1-u_1^2|\tanh^2{r_k}-|1-u_1^2|}\\&+\frac{\frac{1}{2}(1+u_2\tanh{r_k})|1-u_1^2|'\tanh{r_k}}{|1-u_1^2|\tanh^2{r_k}-|1-u_1^2|}\\&-\frac{u_2'|1-u_1^2|\tanh^2{r_k}}{|1-u_1^2|\tanh^2{r_k}-|1-u_1^2|} ,\end{split}
\end{equation}
\begin{equation}    
\begin{split}       
\phi'_k=&-\frac{1}{2}(k+kc^2_s+\frac{a^2V_{\phi\phi}}{k})+\frac{1}{2}[(\frac{a^2V_{\phi\phi}}{2k}-\frac{k}{2}+\frac{kc^2_s}{2})\cos{2\phi_k}+\frac{z'}{z}\sin{2\phi_k}]\\&(|1-u_1^2|^{-\frac{1}{2}}(u_2+\coth{r_k})+|1-u_1^2|^{\frac{1}{2}}\frac{\tanh{r_k}}{1+u_2\tanh{r_k}}),
\end{split}
\end{equation}
where it is corresponding to Eq. \eqref{r prime} and \eqref{phik prime}. 

\section{The calculation of RDM for OTMSS}
\label{The calculation of RDM for OTMS}
In this appendix, we provide the detailed calculation of the Reduced Density Matrix (RDM) for the OTMSS \eqref{normalized otmss}. First, we define the corresponding density operator as follows: 
\begin{equation}
\rho=\sum_{i,j=0}^{\infty}p_{i,j}\ket{\mathcal{O}_i}\bra{\mathcal{O}_j}
\label{total rho}
\end{equation}
where $p_{i,j}$ represents the probability coefficients for the RDM and $\mathcal{O}_i$ denotes the normalized OTMSS as defined in Eq. \eqref{normalized otmss}. By substituting this wave function into the expression for the total density matrix \eqref{total rho}, we arrive at the following result:  
\begin{equation}
\begin{split}
 \rho=&\sum_{n=0}^{\infty}\sum_{m=0}^{\infty}(-1)^n(-1)^m e^{2i\phi_kn} e^{-2i\phi_pm}\\&\frac{\sqrt{(1+u_2\tanh{r_k})^2-|1-u_1^2|\tanh^2{r_k}}}{1+u_2\tanh{r_k}}\frac{\sqrt{(1+u_2\tanh{r_k})^2-|1-u_1^2|\tanh^2{r_k}}}{1+u_2\tanh{r_p}}|1-u_1^2|^{\frac{n}{2}}\\&|1-u_1^2|^{\frac{m}{2}}\frac{ \tanh^n{r_k}}{(1+ u_2\tanh{r_k})^n}\frac{ \tanh^m{r_k}}{(1+ u_2\tanh{r_k})^m}\ket{n_k,n_{-k}}\bra{m_p,m_{-p}},
 \end{split}
 \label{rho1}
\end{equation}
To facilitate the numerical implementation, we define the probability coefficients for the $n$-th and $m$-th components as follows: 
\begin{equation}
\begin{split}
 p_n=(-1)^n e^{-2i\phi_kn}\frac{\sqrt{(1+u_2\tanh{r_k})^2-|1-u_1^2|\tanh^2{r_k}}}{1+u_2\tanh{r_k}}|1-u_1^2|^{\frac{n}{2}}\frac{ \tanh^n{r_k}}{(1+ u_2\tanh{r_k})^n},\\p_m=(-1)^m e^{-2i\phi_pm}\frac{\sqrt{(1+u_2\tanh{r_k})^2-|1-u_1^2|\tanh^2{r_k}}}{1+u_2\tanh{r_p}}|1-u_1^2|^{\frac{m}{2}}\frac{ \tanh^m{r_p}}{(1+ u_2\tanh{r_p})^m}. 
 \end{split}
 \label{pn and pm}
\end{equation}
Based on this notation, we could derive the RDM of OTMSS as follows,
\begin{equation}
\begin{split}
    \rho_k&=Tr_{-k}(\rho)=\sum_{l=0}^\infty \bra{l_{-k}}\rho\ket{l_{-k}}\\\\&=\sum_{l=0}^\infty\sum_{n=0}^{\infty}\sum_{m=0}^{\infty}p_np_m\bra{l_{-k}}\ket{n_k,n_{-k}}\bra{m_p,m_{-p}}\ket{l_{-k}}\\\\&=\sum_{l=0}^\infty\sum_{n=0}^{\infty}\sum_{m=0}^{\infty} p_np_m(I\otimes\bra{l_{-k}})(\ket{n_k}\otimes\ket{n_{-k}})(\bra{m_p}\otimes\bra{m_{-p}})(I\otimes\ket{l_{-k}})\\\\&=\sum_{n=0}^{\infty}\frac{(1+u_2\tanh{r_k})^2-|1-u_1^2|\tanh^2{r_k}}{(1+u_2\tanh{r_k})^2}|1-u_1^2|^{n}\frac{ \tanh^{2n}{r_k}}{(1+ u_2\tanh{r_k})^{2n}}\ket{n_k}\bra{n_k}.
    \label{Rdm}
\end{split}
\end{equation}

\section{The calculation of Purity, von Neumann Entropy, Rényi Entropy, and Logarithmic Negativity}
\label{Purity,von Neumann Entropy, Rényi Entropy, and Logarithmic Negativity}
In this appendix, we provide the detailed derivations for purity, von Neumann entropy, Rényi entropy, and logarithmic negativity based on the RDM defined in Eq. \eqref{Rdm}. We first consider the purity \eqref{purity}, the calculation of which is as follows:
\begin{equation}
    \begin{split}
u=&Tr(\rho^2_k)\\\\=&Tr(\sum_{n=0}^{\infty}\frac{(1+u_2\tanh{r_k})^2-|1-u_1^2|\tanh^2{r_k}}{(1+u_2\tanh{r_k)^2}}|1-u_1^2|^{n}\frac{ \tanh^{2n}{r_k}}{(1+ u_2\tanh{r_k})^{2n}}\\\\&\frac{(1+u_2\tanh{r_k})^2-|1-u_1^2|\tanh^2{r_k}}{(1+u_2\tanh{r_k})^2}|1-u_1^2|^{m}\frac{ \tanh^{2m}{r_k}}{(1+ u_2\tanh{r_k})^{2m}}\ket{n_k}\bra{n_k}\ket{m_k}\bra{m_k})\\\\=&Tr(\sum_{n=0}^{\infty}\frac{(1+u_2\tanh{r_k})^2-|1-u_1^2|\tanh^2{r_k}}{(1+u_2\tanh{r_k})^2}|1-u_1^2|^{n}\frac{ \tanh^{2n}{r_k}}{(1+ u_2\tanh{r_k})^{2n}}\\\\&\frac{(1+u_2\tanh{r_k})^2-|1-u_1^2|\tanh^2{r_k}}{(1+u_2\tanh{r_k})^2}|1-u_1^2|^{m}\frac{ \tanh^{2m}{r_k}}{(1+ u_2\tanh{r_k})^{2m}}\ket{n_k}\delta_{nm}\bra{m_k})\\\\=&\frac{(1+u_2\tanh{r_k})^2-|1-u_1^2|\tanh^2{r_k}}{(1+u_2\tanh{r_k})^2+|1-u_1^2|\tanh^2{r_k}}.
    \end{split}
\end{equation}
Then, it is the von Neumann entropy \eqref{van neumann entropy} as follows,
\begin{equation}
    \begin{split}
        S(\rho_k)=&-Tr(\rho_k\ln{\rho_k})\\=&-Tr[\rho_k\ln{(\sum_{n=0}^{\infty}  P_n)\ket{n_k}\bra{n_k}}]\\=&-Tr[\sum_{n=0}^{\infty} P_n\ln{(\sum_{n=0}^{\infty}P_n)}\ket{n_k}\bra{n_k}\ket{n_k}\bra{n_k}]\\=&-Tr[\sum_{n=0}^{\infty} P_n\ln{(\sum_{n=0}^{\infty}P_n)}\ket{n_k}\bra{n_k}]\\=&-\ln\!\left(1-\frac{|1-u_1^2|\tanh^{2}{r_k}}{(1+u_2\tanh{r_k})^{2}}\right)
-\frac{\dfrac{|1-u_1^2|\tanh^{2}{r_k}}{(1+u_2\tanh{r_k})^{2}}}
{1-\dfrac{|1-u_1^2|\tanh^{2}{r_k}}{(1+u_2\tanh{r_k})^{2}}}
\ln\!\left(\frac{|1-u_1^2|\tanh^{2}{r_k}}{(1+u_2\tanh{r_k})^{2}}\right). 
    \end{split}
\end{equation}
 Next, we calculate the Rényi Entropy \eqref{Smu} as follows, 
 \begin{equation}
    \begin{split}
        S_{\mu}(r_k)&=\frac{1}{1-\mu}\ln{(\sum_{n=0}^{\infty}P_n^u)}
        \\\\&=\frac{1}{1-\mu}\ln{(\sum_{n=0}^{\infty}\frac{((1+u_2\tanh{r_k})^2-|1-u_1^2|\tanh^2{r_k})^\mu}{(1+u_2\tanh{r_k})^{2\mu}}|1-u_1^2|^{\mu n}\frac{ \tanh^{2\mu n}{r_k}}{(1+ u_2\tanh{r_k})^{2\mu n}} )}\\\\&=\frac{1}{1-\mu}[\ln{(\frac{((1+u_2\tanh{r_k})^2-|1-u_1^2|\tanh^2{r_k})^\mu}{(1+u_2\tanh{r_k})^{2\mu}})}+\ln{(\sum_{n=0}^{\infty}\frac{ |1-u_1^2|^{u}\tanh^{2\mu}{r_k}}{(1+ u_2\tanh{r_k})^{2\mu}} )^n}]\\\\&=\frac{1}{1-\mu}[\mu(\ln{(1-\frac{|1-u_1^2|\tanh^2{r_k}}{(1+u_2\tanh{r_k})^2})}+\ln{\frac{|1-u_1^2|^{\mu}\tanh^{2\mu}{r_k}}{(1+ u_2\tanh{r_k})^{2\mu}-|1-u_1^2|^{\mu}\tanh^{2\mu}{r_k}}}]\\\\&=\frac{1}{1-\mu}[\mu\ln{(1-\frac{|1-u_1^2|\tanh^2{r_k}}{(1+u_2\tanh{r_k})^2})}-\ln{(1-\frac{|1-u_1^2|^\mu\tanh^{2\mu}{r_k}}{(1+u_2\tanh{r_k})^{2\mu}})}]
    \end{split}
    \label{renyi entropy}
\end{equation}
where $\mu$ is the Rényi parameter and $P_n$ are the eigenvalues of the RDM for the squeezed vacuum state. The Rényi entropy possesses two noteworthy special cases: $S_{2}(r_k)$ and $S_{1/2}(r_k)$, the formulas for which are derived as follows:
\begin{equation}    
\begin{split}       
S_2(r_k)&=-2\ln{(1-\frac{|1-u_1^2|\tanh^2{r_k}}{(1+u_2\tanh{r_k})^2})}+\ln{(1-\frac{|1-u_1^2|^2\tanh^{4}{r_k}}{(1+u_2\tanh{r_k})^{4}})}\\&=\ln{\frac{1+\frac{|1-u_1^2|\tanh^2{r_k}}{(1+u_2\tanh{r_k})^2}}{1-\frac{|1-u_1^2|\tanh^2{r_k}}{(1+u_2\tanh{r_k})^2}}} ,   
\end{split}
\label{S2}
\end{equation}
\begin{equation}    
\begin{split}        
S_{1/2}(r_k)&=\ln{(1-\frac{|1-u_1^2|\tanh^2{r_k}}{(1+u_2\tanh{r_k})^2})}-2\ln{(1-\frac{|1-u_1^2|^{\frac{1}{2}}\tanh{r_k}}{1+u_2\tanh{r_k}})}]\\&=\ln{[(1+\frac{|1-u_1^2|^{\frac{1}{2}}\tanh{r_k}}{1+u_2\tanh{r_k}})/(1-\frac{|1-u_1^2|^{\frac{1}{2}}\tanh{r_k}}{1+u_2\tanh{r_k}})]}.    
\end{split}
\label{S12}
\end{equation}
where Eqs. \eqref{S2} and \eqref{S12} correspond to the specific cases in Eqs. \eqref{S20} and \eqref{S120}, respectively. Finally, we provide the derivation for the logarithmic negativity \eqref{negativity}, which is initially defined as:
\begin{equation}
     \begin{split}
          E_{\mathcal N}=\log||\rho^{T_{-k}}||_1,
     \end{split}
     \label{EN}
 \end{equation}
where 
\begin{equation}
    \begin{split}
       ||\rho^{T_{-k}}||_1=&\sum_{n=0}^{\infty}|\lambda_n|+\sum_{a<b}^{\infty}2|\lambda_{\pm}|\\\\=&\frac{(1+u_2\tanh{r_k})^2-|1-u_1^2|\tanh^2{r_k}}{(1+u_2\tanh{r_k})^2}[\sum_{n=0}^{\infty}(\frac{ (\tanh^2{r_k}|1-u_1^2|)}{(1+ u_2\tanh{r_k})^2})^n+2\sum_{a<b}^{\infty}(\frac{ |1-u_1^2|^{\frac{1}{2}}\tanh{r_k}}{(1+ u_2\tanh{r_k})})^{a+b}]\\\\=&1+2\frac{(1+u_2\tanh{r_k})^2-|1-u_1^2|\tanh^2{r_k}}{(1+u_2\tanh{r_k})^2}\frac{\frac{ |1-u_1^2|^{\frac{1}{2}}\tanh{r_k}}{(1+ u_2\tanh{r_k})}}{(1+(\frac{ |1-u_1^2|^{\frac{1}{2}}\tanh{r_k}}{(1+ u_2\tanh{r_k})})^2)(1-\frac{ |1-u_1^2|^{\frac{1}{2}}\tanh{r_k}}{(1+ u_2\tanh{r_k})})}\\=&\frac{[(1+u_2\tanh{r_k})^2+|1-u_1^2|\tanh^2{r_k}][(1+u_2\tanh{r_k})-|1-u_1^{\frac{1}{2}}|\tanh{r_k}]}{[(1+u_2\tanh{r_k})^2+|1-u_1^2|\tanh^2{r_k}][(1+u_2\tanh{r_k})-|1-u_1^{\frac{1}{2}}|\tanh{r_k}]}\\&+\frac{2(1+u_2\tanh{r_k})^2|1-u_1^2|^{\frac{1}{2}}\tanh^2{r_k}}{[(1+u_2\tanh{r_k})^2+|1-u_1^2|\tanh^2{r_k}][(1+u_2\tanh{r_k})-|1-u_1^{\frac{1}{2}}|\tanh{r_k}]}. 
     \end{split}
\end{equation}
Thus, the analytical expression for the logarithmic negativity \eqref{negativity} can be derived as follows:
\begin{equation}
    \begin{split}
        E_{\mathcal N}=\log{\frac{(1+ u_2\tanh{r_k}+\tanh{r_k}|1-u_1^2|^{\frac{1}{2}})}{(1+ u_2\tanh{r_k}-\tanh{r_k}|1-u_1^2|^{\frac{1}{2}})}}. 
    \end{split}
\end{equation}




\bibliographystyle{unsrt}
\bibliography{reference}
\end{document}